\documentclass[a4paper,aps,preprintnumbers,showpacs,superscriptaddress,nofootinbib,amsmath,amssymb]{revtex4-1}
\usepackage[dvips]{graphicx}

\begin{document}
\title{Stability of the rotating string loops in Kerr spacetime}
\author{Mariia Churilova}\email{wwrttye@gmail.com}
\affiliation{Research Centre for Theoretical Physics and Astrophysics, Institute of Physics, Silesian University in Opava, CZ-74601 Opava, Czech Republic}
\author{Zden\v{e}k Stuchl{\'i}k}\email{zdenek.stuchlik@physics.slu.cz}
\affiliation{Research Centre for Theoretical Physics and Astrophysics, Institute of Physics, Silesian University in Opava, CZ-74601 Opava, Czech Republic}

\begin{abstract}
We study stability of the circular string loops rotating in the equatorial plane of Kerr spacetime against equatorial and polar perturbations. We consider motion of such string loops for different modes in the case of arbitrary equation of state. We also obtain analytical expression for the fundamental frequencies of the string loop oscillations under equatorial and polar perturbations.
\end{abstract}

\maketitle

\section{Introduction}

Relativistic elastic strings, that is, one-dimensional elastic bodies whose internal energy depends only on their stretching, were first studied by Carter \cite{Carter89, Carter89b} as models for superconducting cosmic strings. Relativistic elasticity has been intensely developed and has a wide range of applications (see \cite{Nat-Que-Leo:2018:CQG:} and references therein). In particular, formal equivalence of the equations for relativistic string dynamics and the relativistic magnetohydrodynamics equations was established in \cite{SemBer}.

We consider equilibrium configuration of the string in different black hole spacetimes. We study stability of the string's equilibrium configuration under various perturbations for arbitrary equation of state of the string. In addition to the altogether stability of the string's equilibrium configuration we also study its stability for different modes of the perturbations.
Stability of the strings in black hole spacetimes has been previously studied in \cite{Carter89b,FS91,LS93,Larsen94a,Larsen94b,BS09,Nat-Que-Leo:2018:CQG:,Churilova:2023swu}.

In this paper we extend results of \cite{Nat-Que-Leo:2018:CQG:} and \cite{Churilova:2023swu} on stability of the rotating elastic string loops from Schwarzschild to Kerr spacetime. Namely, we consider axially symmetric string loops rotating in the equatorial plane of the Kerr spacetime and study stability of their equilibrium configuration under equatorial and polar perturbations. We establish connection between stability issues and existing of the real frequencies of the string loop oscillations under equatorial and polar perturbations and for fundamental mode we present analytical expression for the frequencies.

The paper is organized as follows. In Section II we give a brief derivation of the equations of motion (see \cite{Nat-Que-Leo:2018:CQG:} and references therein) for the reader's convenience. In Section III we re-derive equilibrium conditions for an axially symmetric elastic string loop rotating in the equatorial plane of Kerr spacetime found in \cite{Nat-Que-Leo:2018:CQG:}. In Section IV we obtain linearized equations of motion under equatorial and polar perturbations in Kerr spacetime following the scheme used in \cite{Nat-Que-Leo:2018:CQG:}. In Section V we find analytical expression for the fundamental frequencies of the string loop oscillations under equatorial and polar perturbations, extending corresponding results of \cite{Churilova:2023swu}. Section VI is devoted to stability analysis of the rotating string loops equilibria in Kerr spacetime and establishing connections between stability and existing of the real frequencies of the string loop oscillations under equatorial and polar perturbations extending results of \cite{Churilova:2023swu}. Conclusions contain summary of the results.

\section{Equations of motion}

The string's worldsheet is described by the spacetime coordinates $X^{\alpha}(\tau,\lambda)$ with $\alpha = 0,1,2,3$, given as functions of two worldsheet coordinates: $\tau$ parameterizes string evolution, $\lambda$ is running along the string. Here the parameter $\lambda$ is the arclength in the string's unstretched configuration. The embedding $X$ induces a metric
\begin{equation}\label{metric}
h_{AB} = g_{\mu\nu}(X) \partial_A X^\mu \partial_B X^\nu
\end{equation}
on the worldsheet, where $ g_{\mu\nu}$ is the background metric and $A,B\in\left\{\tau,\lambda\right\}$.
For an action
\begin{equation}
 S = \int \mathcal{L} \, \mathrm{d} \tau \mathrm{d} \lambda
\end{equation}
of the elastic string with the internal energy density $\rho$ depending only on its stretching,
the Lagrangian density is
\begin{equation} \label{Lagrangian}
{\mathcal{L}} = \rho \sqrt{-h},
\end{equation}
where $h \equiv \det(h_{AB})$.
Then the variation $\delta {\mathcal{L}}$ of the Lagrangian density resulting from a variation $\delta X$ of the embedding has the form
\begin{equation}
\delta {\mathcal{L}} = - \frac12 \sqrt{-h} \, T^{AB} \delta h_{AB},
\end{equation}
where the string's energy-momentum tensor $T^{AB}$  is given by the relation
\begin{equation} \label{TAB}
T^{AB} = \left(p+\rho\right) \frac{-\delta^A_{\tau} \delta^B_{\tau}}{h_{\tau \tau}} + p h^{AB}\,\,
\end{equation}
with $p$ the string's pressure, $\rho$ the string's energy density.
Due to Hamilton's principle
\begin{equation}
\delta \int {\mathcal{L}} \, d\tau d\lambda = 0,
\end{equation}
we obtain the equations of motion in the form
\begin{equation} \label{motion}
\frac1{\sqrt{-h}}\partial_B \left( \sqrt{-h} \, T^{AB}\partial_A X^\alpha \right) + T^{AB} \Gamma^\alpha_{\mu\nu} \partial_A X^\mu \partial_B X^\nu = 0\,\,.
\end{equation}

Taking the Kerr metric as a background metric $ g_{\mu\nu}$, we find the induced metric (\ref{metric}) and the energy-momentum tensor (\ref{TAB}) using the embedding describing the string's equilibrium state for arbitrary string's pressure $p$ and string's energy density $\rho$. Substituting into the equations of motion (\ref{motion}), we obtain the condition for the equilibrium configuration. We then change the embedding to that describing the oscillatory state of the string and finding the induced metric (\ref{metric}) and the energy-momentum tensor (\ref{TAB}) to first order, we obtain the linearized equations of motion for the equatorial and polar oscillations of the string satisfying equilibrium condition.

\section{Equilibrium conditions}

We consider an axially symmetric elastic string loop rotating in the equatorial plane with a constant angular velocity $\Omega$ in equilibrium at a constant radius R, described by an embedding
\begin{equation} \label{equilibrium}
	\begin{cases}
		t(\tau,\lambda)=\tau \\
		r(\tau,\lambda) = R \\
		\theta(\tau, \lambda)=\frac{\pi}{2} \\
		\varphi(\tau,\lambda) = \Omega \tau +  k \lambda \, ,
	\end{cases}
\end{equation}
where $k$ is a constant and $\lambda \in \left[0,\frac{2\pi}{k}\right]$. Due to the definition of $\lambda$,
\begin{equation}
k = \frac1R_0  \,\, ,
\end{equation}
where $R_0$ is the radius of the unstretched string loop in flat spacetime. Taking the Kerr solution with mass $M\geq 0$ and angular momentum $Ma$
\begin{equation}\label{Kerr}
ds^2 = - \left( 1 - \frac{2Mr}{\Delta^2} \right) dt^2 + \frac{\Delta^2}{\Sigma} dr^2 + \Delta^2 d \theta^2 + \left(r^2 + a^2 + \frac{2Mra^2}{\Delta^2} \sin^2 \theta\right) \sin^2 \theta d\varphi^2 - \frac{4Mra\sin^2 \theta}{\Delta^2} dt d\varphi,
\end{equation}
where $\Delta^2 = r^2 + a^2 \cos^2 \theta$ and $\Sigma = r^2-2Mr+a^2$, as a background metric $ g_{\mu\nu}$, we obtain from (\ref{metric}) induced metric as
\begin{equation}\label{induced equilibrium}
\left(h_{AB}\right) =
\left(
\begin{array}{cc}
 \frac{2 M (a \Omega -1)^2+R \left(a^2 \Omega ^2+R^2 \Omega ^2-1\right)}{R}
   & \frac{\Omega  R^3-2 a M+a^2 (2 M+R) \Omega }{R} \\
 \frac{\Omega  R^3-2 a M+a^2 (2 M+R) \Omega }{R} & \frac{(2 M+R) a^2}{R}+R^2
   \\
\end{array}
\right)\;.
\end{equation}
Substituting into (\ref{TAB}) and (\ref{motion}), we obtain from the equations of motion the condition for equilibrium configurations (\ref{equilibrium}) in the form
\begin{equation} \label{Kerr_eq}
\frac{p R^2 (M-R)}{a^2+R^2-2MR}=\frac{(p+\rho ) \left(M (a \Omega-1)^2-R^3 \Omega^2\right)}{a^2 \Omega^2+R^2\Omega^2-1+\frac{2 M}{R} (a \Omega-1)^2}\,\,.
\end{equation}
Equilibrium condition (\ref{Kerr_eq}) allows one to express angular velocity $\Omega$ in terms of the string's pressure $p$ and the string's energy density $\rho$ or, alternatively, in terms of the transverse
\begin{equation}
s = \sqrt{-\frac{p}{\rho}}\,\,
\end{equation}
and longitudinal
\begin{equation} \label{c^2}
c = \sqrt{\frac{dp}{d\rho}}\,\,
\end{equation}
speeds of sound.

\begin{figure}[h]
	\centering 	
	\includegraphics[scale=0.9]{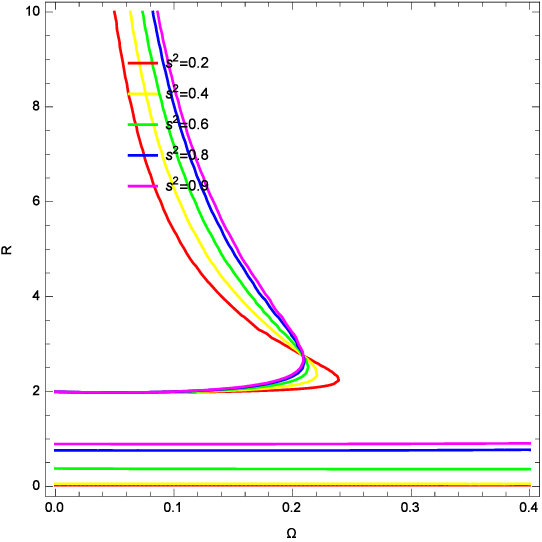}\includegraphics[scale=0.9]{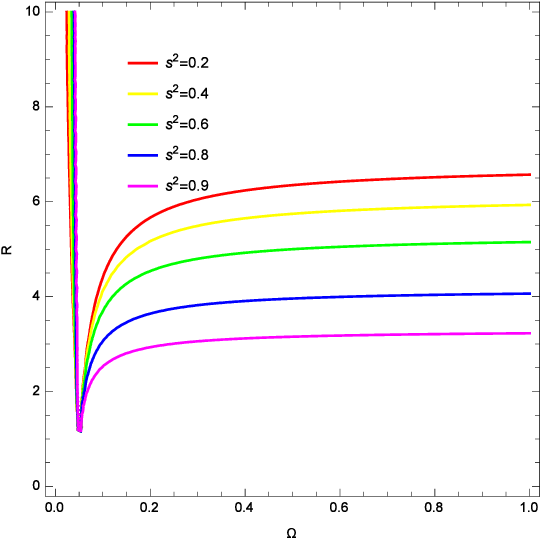}
\caption{\label{a}
{\large Equilibrium radius as a function of angular velocity for different values of the transverse speed of sound, $M=1$: left panel $a=0.2$, right panel $a=20$.}}
\end{figure}

Equilibrium radius R, satisfying equilibrium condition (\ref{Kerr_eq}), is depicted on FIG. \ref{a} as a function of the angular velocity $\Omega$ for different equations of state characterized by the values of the transverse speed of sound $s$. Left panel shows that for the rotating black hole with $a=0.2$ equilibrium radius is defined unambiguously outside the retrograde photon sphere, which is a natural requirement for the string loop to fulfill. For all equations of state equilibrium radius R is decreasing with increasing of the angular velocity $\Omega$ of the string loop. Moreover, the larger the transverse speed of sound, the larger the equilibrium radius for the same fixed angular velocity. On the contrary, right panel for $a=20$ shows that this behaviour is changed dramatically after $\Omega=0.5$: the equilibrium radius is increasing with increasing of the angular velocity of the string loop and the larger transverse speed of sound corresponds to the less equilibrium radius for the same fixed angular velocity.

\section{Linear stability}

In order to analyse the linear stability of the string loop rotating in the equatorial plane at the equilibrium configuration (\ref{equilibrium}), that is, satisfying equilibrium condition (\ref{Kerr_eq}), we consider the equatorial perturbations $\delta r$ and $\delta \varphi$ and the polar perturbations $\delta \theta$. Substituting the embedding
\begin{equation} \label{embed2}
	\begin{cases}
		t(\tau,\lambda)=\tau \\
		r(\tau,\lambda) = R + \delta r(\tau, \lambda)\\
		\theta(\tau, \lambda)=\frac{\pi}{2}+\delta \theta(\tau, \lambda) \\
		\varphi(\tau,\lambda) = \Omega \tau +  k \lambda + \delta \varphi(\tau,\lambda)
	\end{cases}
\end{equation}
and Kerr metric (\ref{Kerr}) as a background metric $ g_{\mu\nu}$
into equation (\ref{metric}), we obtain induced metric to the first order as
\begin{equation} \label{induced linearized}
\begin{array}{ll}
h_{\tau\tau}=
 a^2 \Omega ^2+R^2 \Omega ^2+\frac{2 a^2 M \Omega ^2}{R}-\frac{4 a M \Omega
   }{R}+\left(2 R \Omega ^2-\frac{2 a^2 M \Omega ^2}{R^2}+\frac{4 a M \Omega
   }{R^2}-\frac{2 M}{R^2}\right) \delta r\\
   \hspace{20pt} +\left(\frac{4 M \Omega  a^2}{R}+2 \Omega  a^2-\frac{4 M a}{R}+2 R^2
   \Omega \right) \delta \dot{\varphi}+\frac{2 M}{R}-1\;,\\
 h_{\tau\lambda}= h_{\lambda\tau}= k
   \Omega  a^2+\frac{2 k M \Omega  a^2}{R}-\frac{2 k M a}{R}+k R^2 \Omega
   +\left(-\frac{2 k M \Omega  a^2}{R^2}+\frac{2 k M a}{R^2}+2 k R \Omega
   \right) \delta r\\
   \hspace{20pt}+\left(\frac{2 M \Omega
   a^2}{R}+\Omega  a^2-\frac{2 M a}{R}+R^2 \Omega \right) \delta \varphi'+\left(k a^2+\frac{2 k M a^2}{R}+k R^2\right)
   \delta \dot{\varphi} \;,\\
 h_{\lambda\lambda}=a^2 k^2+R^2 k^2+\frac{2 a^2 M
   k^2}{R}+\left(2 k^2 R-\frac{2 a^2 k^2 M}{R^2}\right) \delta r+\left(2 k a^2+\frac{4 k M a^2}{R}+2 k R^2\right)
   \delta \varphi' \;,\\
\end{array}
\end{equation}
where $\,\,\dot{} \equiv \frac{\partial}{\partial \tau}$ and $\,\,'\equiv \frac{\partial}{\partial \lambda}$.
Substituting into (\ref{TAB}) and (\ref{motion}), we obtain the linearized equations of motion with equatorial and polar perturbations decoupled
\begin{equation}\label{le1}
 A \, \delta r'+B \, \delta \varphi ''+ C \, \delta \dot{r}+ D \, \delta \dot{\varphi}' + E \, \delta \ddot{\varphi}=0 \,\, ,
\end{equation}
\begin{equation}\label{le2}
 F \, \delta r + G \, \delta \varphi'+H \, \delta r''+I \, \delta \dot{\varphi}+ J \, \delta \dot{r}'+L \, \delta \ddot{r}=0 \,\, ,
\end{equation}
\begin{equation} \label{le3}
 N\, \delta \theta + H\, \delta \theta''+ J\, \delta \dot{\theta}'+ L\, \delta \ddot{\theta} =0 \,\, ,
\end{equation}
where the coefficients depend on $M$, $a$, $R$, $s$ and $c$ as given in Appendix A.

Fourier expanding the perturbations and
looking for solutions proportional to $e^{i \omega \tau}$, we arrive at the characteristic polynomials: for the equatorial perturbations
\begin{equation}
\begin{split} \label{polynomialequ}
\tilde{p}_j(\omega)
&= \frac{1}{E L}\left[ k^2 j^2 ( A G -B F + B H k^2 j^2) +
k j \left( C G + A I -D F + (D H + B J) k^2 j^2 \right) \omega  \right. \\ &\left. +\left(  C I -E F +
k^2 (E H + D J + B L) j^2 \right) \omega^2  +
k (E J + D L) j \, \omega^3 + E L\, \omega^4\right]
\end{split}
\end{equation}
and for the polar perturbations
\begin{equation} \label{polynomialpol}
\tilde{q}_j(\omega)=L\, \omega^2+j k J\, \omega -\left( N -j^2 k^2 H \right) \,\, ,
\end{equation}
from which we can conclude about linear stability of the string loop under equatorial (polar) perturbations for each mode $j \in \mathbb{Z}$ only if all the roots of $\tilde{p}_j$ ($\tilde{q}_j$) have a non-negative imaginary part. Since complex roots come in conjugate pairs, this necessary condition amounts to requiring all the roots to be real.

\section{Fundamental frequencies}

Considering characteristic polynomials for the equatorial (\ref{polynomialequ}) and polar (\ref{polynomialpol}) perturbations for each mode $j \in \mathbb{Z}$, we obtain the frequencies $\omega$ of the string loop oscillations under equatorial (polar) perturbations as the roots of the corresponding characteristic polynomials (\ref{polynomialequ}) and (\ref{polynomialpol}). For the zero mode $j=0$
characteristic polynomial for the equatorial perturbations takes the form
\begin{equation}\label{polynomialequZeroMode}
 \tilde{p}_0(\omega)=\left(C I -E F\right) \omega^2  + E L \, \omega^4
\end{equation}
and for the polar perturbations the form
\begin{equation} \label{polynomialpolZeroMode}
\tilde{q}_0(\omega)=L\, \omega^2- N.
\end{equation}
Thus fundamental frequencies of the string loop oscillations under equatorial (polar) perturbations being the roots of polynomials (\ref{polynomialequZeroMode}) and (\ref{polynomialpolZeroMode}) are found as
\begin{align*}
 & \omega_r^2=\omega_\varphi^2=\frac{\left(E F -C I\right)}{E L},
\end{align*}

\begin{align*}
 & \omega_\theta^2=\frac{N}{ L}
\end{align*}
with coefficients $E$, $F$, $C$, $I$, $L$, $N$ given in Appendix A.

\begin{figure}[h]
	\centering 	
	\includegraphics[scale=0.9]{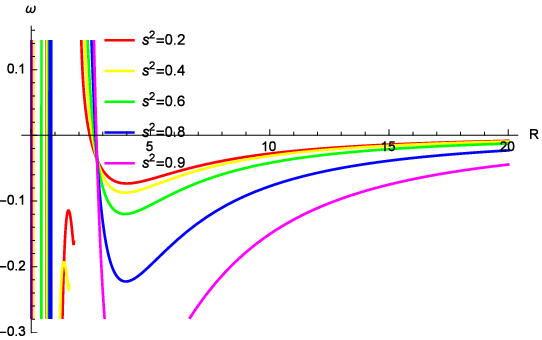}\includegraphics[scale=0.9]{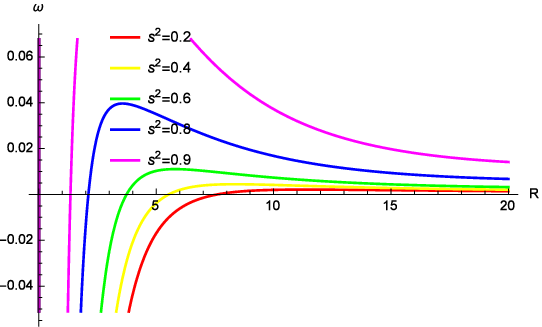}
\caption{\label{RadialEq}
{\large Radial profiles of the fundamental frequency $ \omega_r=\omega_\varphi$ for different values of the transverse speed of sound, $M=1$: left panel $a=0.2$, right panel $a=20$.}}
\end{figure}

\begin{figure}[h]
	\centering 	
	\includegraphics[scale=0.9]{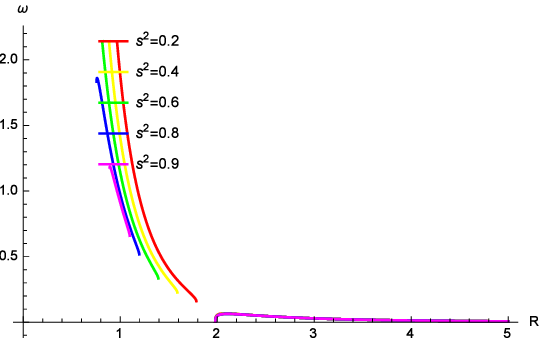}\includegraphics[scale=0.9]{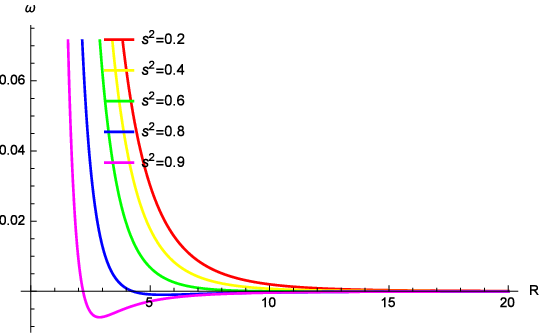}
\caption{\label{RadialPol}
{\large Radial profiles of the fundamental frequency $\omega_\theta$ for different values of the transverse speed of sound, $M=1$: left panel $a=0.2$, right panel $a=20$.}}
\end{figure}

Equating the fundamental frequencies to zero $\omega_r^2=\omega_\varphi^2=0$, $\omega_\theta^2=0$, we can find marginally stable positions of the string loop rotating in the equatorial plane. Figs. \ref{RadialEq}-\ref{RadialPol} show radial profiles of the fundamental frequencies for different values of the transverse speed of sound $s^2$ for the equatorial and polar perturbations. While for the equatorial perturbations the character of changing at $a=0.2$ and $a=20$ is opposite both in radius and transverse speed of sound, for the polar perturbations the direction of changing at $a=0.2$ and $a=20$ is the same.

\section{Stability analysis}

Stability of the string loop under equatorial (polar) perturbations for each mode $j \in \mathbb{Z}$ requires all the roots of the characteristic polynomial $\tilde{p}_j$ ($\tilde{q}_j$) to be real.

For the polar perturbations characteristic polynomial (\ref{polynomialpol}) is a quadratic one on the variable $\omega$. Since in Schwarzschild's spacetime the discriminant of this polynomial is positive (see \cite{Nat-Que-Leo:2018:CQG:}), the polynomial (\ref{polynomialpol}) has two distinct real roots. This means that the equilibrium of the rotating string loop in Schwarzschild's spacetime is linearly stable under polar perturbations for all the modes $j \in \mathbb{Z}$.

This is no longer the case for the rotating string loop in Kerr background, since there exist values of $s^2$ and $a$, for which the discriminant $\Delta_{\tilde{q}_j}$ of the polynomial (\ref{polynomialpol}) is negative, thus the polynomial has two complex conjugate non-real roots. Therefore for these values of $s^2$ and $a$ the rotating string loop in Kerr background is linearly unstable under polar perturbations.

\begin{figure}[h]
	\centering 	
	\includegraphics[scale=0.9]{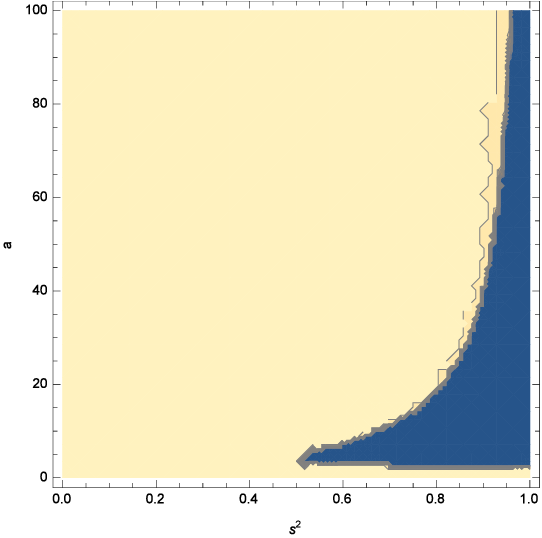}\includegraphics[scale=0.9]{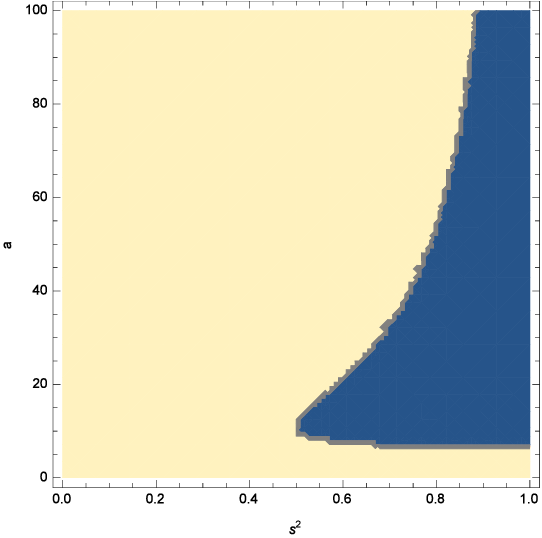}
\caption{\label{Min}
{\large Contour plot of the global minimum of $\text{tanh}\left(\Delta_{\tilde{q}_j}\right)$ as a function of $s^2$ and $a$ for $R=4$, $R=7$ and $R=11$, $M=1$, $k=1$ (yellow - positive values of the global minimum of $\text{tanh}\left(\Delta_{\tilde{q}_j}\right)$, blue - negative ones).}}
\end{figure}

\begin{figure}[h]
	\centering 	
	\includegraphics[scale=0.9]{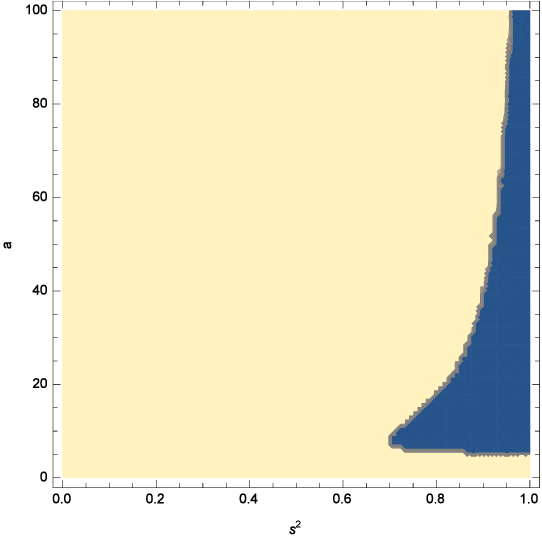}\includegraphics[scale=0.9]{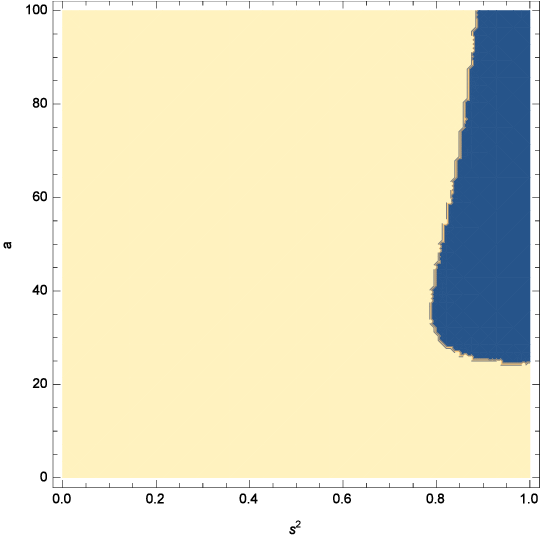}
\caption{\label{1}
{\large Contour plot of $\text{tanh}\left(\Delta_{\tilde{q}_1}\right)$ as a function of $s^2$ and $a$ for $R=4$, $R=7$ and $R=11$, $M=1$, $k=1$ (yellow - positive values of the global minimum of $\text{tanh}\left(\Delta_{\tilde{q}_j}\right)$, blue - negative ones).}}
\end{figure}

\begin{figure}[h]
	\centering 	
	\includegraphics[scale=0.9]{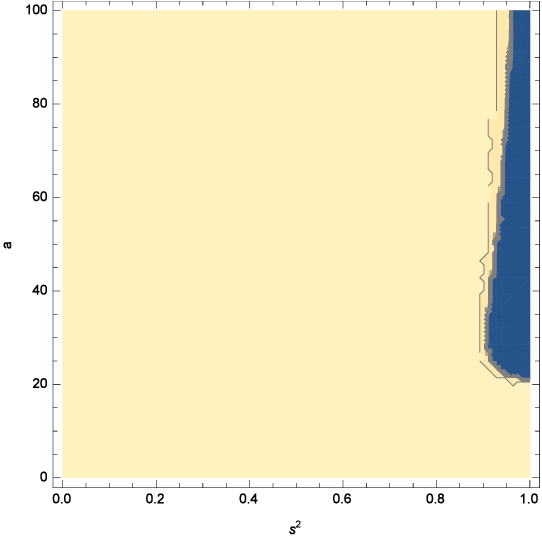}\includegraphics[scale=0.9]{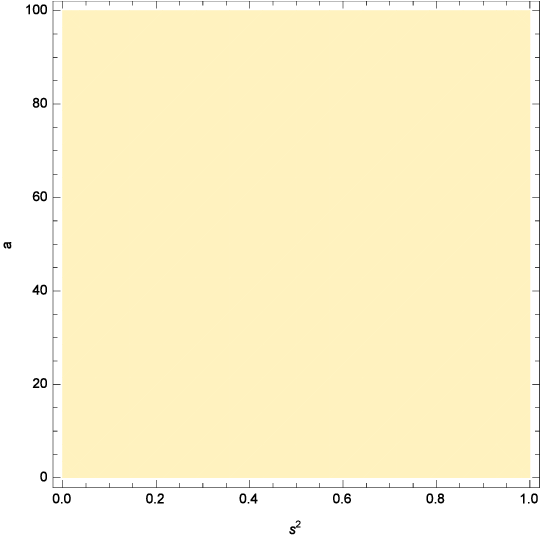}
\caption{\label{4}
{\large Contour plot of $\text{tanh}\left(\Delta_{\tilde{q}_4}\right)$ as a function of $s^2$ and $a$ for $R=4$, $R=7$ and $R=11$, $M=1$, $k=1$ (yellow - positive values of the global minimum of $\text{tanh}\left(\Delta_{\tilde{q}_j}\right)$, blue - negative ones).}}
\end{figure}

The regions of the instability are the largest for the fundamental mode $j=0$ and coincide with the regions of the altogether instability depicted in FIG.\ref{Min}. It can be seen that instability starts for rather small $a$ and for $s^2$ around $0.5$, but when $a$ increases, instability region moves to larger values of $s^2$. With radius $R$ increasing, instability starts for larger values of $a$. For the mode $j=1$ the picture is similar, but the region of instability diminishes considerably (FIG.\ref{1}). For the mode $j=4$ equilibrium is stable up to $a=100$ at $R=11$ already (FIG.\ref{4}), and for the mode $j=30$ it is stable starting from $R=4$.

Thus parameter $a$ causes instability for $s^2$ larger than $0.5$, but for large radius $R$ and higher modes the string loop regains the stability.

For the equatorial perturbations characteristic polynomial (\ref{polynomialequ}) is of degree 4 on the variable $\omega$. Since in Schwarzschild's spacetime the discriminant of this polynomial is negative for the mode $j=1$ (see \cite{Nat-Que-Leo:2018:CQG:}), the polynomial (\ref{polynomialequ}) has two distinct real roots and two complex roots. This means the altogether instability under the equatorial perturbations of the equilibrium of the rotating string loop in Schwarzschild's spacetime.

\begin{figure}[h!]
	\centering 	
\includegraphics{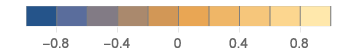}
\caption{\label{Eq0R4}
{\large Contour plot of $\text{tanh}\left(\Delta_{\tilde{p}_0}\right)$ as a function of $s^2$ and $a$ for $R=4$, $M=1$, $k=1$.}}
\end{figure}

On the contrary, in Kerr background the discriminant $\Delta_{\tilde{p}_0}$ of the characteristic polynomial (\ref{polynomialequ}) for the fundamental mode $j=0$ has positive values (FIG.\ref{Eq0R4}), which means that it can (but not necessarily does) have four distinct real roots. This would mean stability for the fundamental mode, so it suffices to find one real root of the characteristic polynomial (\ref{polynomialequ}). Thus finding the roots of the characteristic polynomial (\ref{polynomialequ}) for the values of $a$ and $s^2$ which provide positivity of the discriminant, we can conclude about the linear stability of the rotating string loop in Kerr background under equatorial perturbations. For each mode $j \in \mathbb{Z}$ existence of the real frequency of the rotating string loop oscillations (which is the root of the characteristic polynomial (\ref{polynomialequ})) within the region of the positivity of the discriminant means the string loop stability in that region.

It is worth mentioning that for large values of $a$ and $s^2$ the rotating string loop is linearly unstable against polar perturbations and linearly stable under equatorial perturbations.

\begin{figure}[h]
	\centering 	
	\includegraphics[scale=0.9]{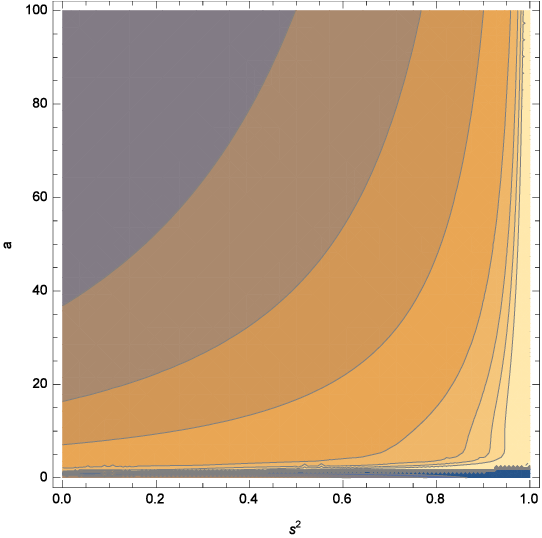}
\end{figure}

\section{Conclusions}

We have studied stability of the circular string loops rotating in the equatorial plane of Kerr spacetime against equatorial and polar perturbations.
Unlike the Schwarzschild's case the equilibrium of the rotating string loop in Kerr spacetime is not linearly stable under polar perturbations for all the modes $j \in \mathbb{Z}$, but has regions of instability caused by parameter $a$ for speed of sound $s^2$ larger than $0.5$, though for large radius $R$ and higher modes the string loop regains stability. Thus although for such values of $a$ and $s^2$ there is no altogether stability, equilibrium is stable for higher modes.

For the equatorial perturbations again unlike the altogether instability under the equatorial perturbations of the equilibrium of the rotating string loop in Schwarzschild's spacetime, in Kerr spacetime there emerge regions of stability in parameters  $a$ and $s^2$ for the fundamental mode already. For each mode $j \in \mathbb{Z}$ existence of the real frequency of the rotating string loop oscillations (which is the root of the characteristic polynomial (\ref{polynomialequ})) within the region of the positivity of the discriminant means the string loop stability in that region. Thus while for large values of $a$ and $s^2$ the rotating string loop is linearly unstable against polar perturbations, it becomes linearly stable under equatorial perturbations.
We also obtained analytical expression for the fundamental frequencies of the string loop oscillations under equatorial and polar perturbations.

\appendix

\section{Linearized equations coefficients} \label{appendixA}


\begin{align*}
& A=\frac{k \left(c^2+s^2\right)}{\eta  \xi }; \\
& B=\frac{R^2 s^2 \left(s^2-1\right) \chi \zeta  }{\eta ^3 k}; \\
& C=2 k R^3; \\
& D=-\frac{2 k^2 R s^2 \zeta  \left(a^5 M R \left(s^2-1\right)-2 a^3
   M R^2 \left(s^2-1\right) (M-2 R)+\psi  \left(a^2 (2
   M+R)+R^3\right)-3 a M R^4 \left(s^2-1\right) (2
   M-R)\right)}{\eta ^2 \xi ^2}; \\
& E=\frac{k R \zeta }{\eta  \chi  \left(a^2+R
   (R-2 M)\right)} \left(a^8 M^2 R \left(s^2-1\right)^2+a^6 M R^2
   \left(2 M^2 \left(s^4+3 s^2-2\right)+M R \left(11 s^4-15
   s^2+2\right)+R^2 \left(s^4-s^2-\right.\right.\right.\\
  & \hspace{40 pt} \left.\left.\left.-2\right)\right)+a^4 R^3
   \left(-4 M^4 \left(2 s^4+s^2-1\right)-4 M^3 R \left(4 s^4-7
   s^2+1\right)+3 M^2 R^2 \left(7 s^4-10 s^2+3\right)+M R^3
   \left(3 s^4-s^2-\right.\right.\right.\\
  & \hspace{40 pt} \left.\left.\left.-4\right)+R^4 \left(s^2+1\right)\right)+a^2 R^6
   \left(M^3 \left(-26 s^4+38 s^2-8\right)+M^2 R \left(9 s^4-19
   s^2+8\right)+3 M R^2 \left(s^4-s^2-2\right)+\right.\right.\\
  & \hspace{40 pt} \left.\left.+2 R^3
   \left(s^2+1\right)\right)+s^2 \psi  \left(2 a^5 M (2 M+R)+4
   a^3 M R^2 (3 M+2 R)+6 a M R^5\right)+R^9 \left(s^2+1\right)
   \left(-2 M^2 \left(s^2-2\right)+\right.\right.\\
  & \hspace{40 pt} \left.\left.+M R
   \left(s^2-4\right)+R^2\right)\right); \\
& F=\frac{\xi  R}{\left(a^2+R (R-2 M)\right)^2 \left(R \left(\Omega ^2
   \left(a^2+R^2\right)-1\right)+2 M (a \Omega -1)^2\right)^2}\left( \left(2 a^4 M \left(s^2-1\right) (M-R)+a^2 R \left(2 M-R\right) \left(s^2 \left(-3 M^2+\right.\right.\right.\right.\\
  & \hspace{40 pt} \left.\left.\left.\left.+M R+R^2\right)+3 M
   (M-R)\right)+\Omega  \left(2 a M \left(-2 a^4 \left(s^2-1\right) (2 M-R)-a^2 R \left(-12 M^2
   \left(s^2-1\right)+M R \left(3 s^2-11\right)+\right.\right.\right.\right.\right.\\
  & \hspace{40 pt} \left.\left.\left.\left.\left.+R^2 \left(s^2+3\right)\right)-R^2 (2 M-R) \left(M^2 \left(8
   s^2-4\right)+\left(s^2-1\right) \left(R^2-3 M R\right)\right)\right)+\Omega  \left(12 a^6 M^2
   \left(s^2-1\right)+a^4 R \left(\left(s^2-\right.\right.\right.\right.\right.\right.\\
  & \hspace{40 pt} \left.\left.\left.\left.\left.\left.-1\right) \left(8 M R^2-36 M^3\right)+\left(s^2+1\right)
   \left(R^3-12 M^2 R\right)\right)+a^2 R^2 \left(24 M^4 \left(2 s^2-1\right)-12 M^3 R
   \left(s^2-1\right)+M^2 R^2 \left(33-\right.\right.\right.\right.\right.\right.\\
  & \hspace{40 pt} \left.\left.\left.\left.\left.\left.-35 s^2\right)+2 M R^3 \left(9 s^2-11\right)-R^4
   \left(s^2-3\right)\right)+\Omega  \left(\Omega  \left(2 a^8 M \left(s^2-1\right) (M+R)-a^6 R \left(6 M^3
   \left(s^2-1\right)+\right.\right.\right.\right.\right.\right.\right.\\
  & \hspace{40 pt} \left.\left.\left.\left.\left.\left.\left.+M^2 R \left(9 s^2-5\right)+M R^2 \left(11-7 s^2\right)+R^3\right)+a^4 R^2 \left(4 M^3
   \left(2 s^2-1\right) (M+R)+M^2 R^2 \left(38-37 s^2\right)+\right.\right.\right.\right.\right.\right.\right.\\
  & \hspace{40 pt} \left.\left.\left.\left.\left.\left.\left.+M R^3 \left(9 s^2-13\right)-2 R^4\right)-a^2
   R^5 \left(M^3 \left(34-38 s^2\right)+M^2 R \left(27 s^2-29\right)-5 M R^2
   \left(s^2-1\right)+R^3\right)-\right.\right.\right.\right.\right.\right.\\
  & \hspace{40 pt} \left.\left.\left.\left.\left.\left.-M R^8 \left(s^2 (M-R)-2 M+R\right)\right)-2 a M \left(2 a^6
   \left(s^2-1\right) (2 M+R)+a^4 R \left(-12 M^2 \left(s^2-1\right)+M R \left(3-11 s^2\right)+\right.\right.\right.\right.\right.\right.\right.\\
  & \hspace{40 pt} \left.\left.\left.\left.\left.\left.\left.+R^2 \left(7
   s^2-11\right)\right)+2 a^2 R^2 \left(M^3 \left(8 s^2-4\right)+M^2 R \left(3 s^2-1\right)-19 M R^2
   \left(s^2-1\right)+R^3 \left(7 s^2-9\right)\right)+\right.\right.\right.\right.\right.\right.\\
  & \hspace{40 pt} \left.\left.\left.\left.\left.\left.+R^5 \left(M^2 \left(38
   s^2-34\right)+\left(s^2-1\right) \left(9 R^2-35 M R\right)\right)\right)\right)+R^5 (2 M-R) \left(M^2
   \left(19 s^2-17\right)+\left(s^2-1\right) \left(2 R^2-\right.\right.\right.\right.\right.\right.\\
  & \hspace{40 pt} \left.\left.\left.\left.\left.\left.-12 M R\right)\right)\right)\right)+M R^2 (R-2 M)^2
   \left(s^2 (2 M-R)-M+R\right)\right)\right); \\
& G=\frac{k R^3 \left(c^2+s^2\right) (M-R)}{\xi  \left(s^2-1\right)}; \\
& H=-\frac{R^5 s^2 \left(s^2-1\right) \chi }{\eta ^2 \xi }; \\
& I=\frac{2 \eta ^2 k^4 R^3 \left(1-s^2\right) }{\xi ^3 \left(s^2-1\right)^2 \chi ^2}\left(\left(2 a^7 M^2 R \left(s^2-1\right)^2-2 a^5 M R^2 \left(M^2
   \left(4 s^4-9 s^2+4\right)+M R \left(-5 s^4+12 s^2-5\right)-R^2 s^2\right)+\right.\right.\\
  & \hspace{40 pt} \left.\left.+2 a^3 M R^3 \left(2 M^3
   \left(2 s^4-5 s^2+2\right)-8 M^2 R \left(2 s^4-5 s^2+2\right)+M R^2 \left(7 s^4-24 s^2+7\right)+4 R^3
   s^2\right)+\right.\right.\\
  & \hspace{40 pt} \left.\left.+\psi  \left(a^4 M \left(M \left(5 s^2-3\right)+R \left(s^2-3\right)\right)+a^2 R \left(M^3
   \left(6-12 s^2\right)+3 M^2 \left(2 R s^2+R\right)+2 M R^2 \left(s^2-3\right)+R^3\right)+\right.\right.\right.\\
  & \hspace{40 pt} \left.\left.\left.+R^4 \left(-3
   M^2 \left(s^2-2\right)+M R \left(s^2-5\right)+R^2\right)\right)+6 a M R^5 \left(2 M^3 \left(2 s^4-5
   s^2+2\right)+M^2 R \left(-4 s^4+13 s^2-4\right)+\right.\right.\right.\\
  & \hspace{40 pt} \left.\left.\left.+M R^2 \left(s^4-6 s^2+1\right)+R^3
   s^2\right)\right)\right); \\
& J=-\frac{2 k^3 R^4 s^2 \left(-\psi  \left(a^2 (2 M+R)+R^3\right)-M R \left(s^2-1\right) \left(a^3+3 a
   R^2\right) \left(a^2+R (R-2 M)\right)\right)}{\eta  \xi ^3}; \\
& L=-\frac{k^4 R^4}{\xi ^3 \chi }\left( \left(a^8 M^2 R \left(s^2-1\right)^2+a^6 M R^2 \left(2 M^2 \left(s^4+3 s^2-2\right)+M R
   \left(11 s^4-15 s^2+2\right)+R^2 \left(s^4-s^2-2\right)\right)+\right.\right.\\
  & \hspace{40 pt}\left. \left.+a^4 R^3 \left(-4 M^4 \left(2
   s^4+s^2-1\right)-4 M^3 R \left(4 s^4-7 s^2+1\right)+3 M^2 R^2 \left(7 s^4-10 s^2+3\right)+M R^3 \left(3
   s^4-s^2-4\right)+\right.\right.\right.\\
  & \hspace{40 pt}\left.\left. \left.+R^4 \left(s^2+1\right)\right)+a^2 R^6 \left(M^3 \left(-26 s^4+38 s^2-8\right)+M^2 R
   \left(9 s^4-19 s^2+8\right)+3 M R^2 \left(s^4-s^2-2\right)+\right.\right.\right.\\
  & \hspace{40 pt}\left.\left. \left.+2 R^3 \left(s^2+1\right)\right)+2 a M s^2
   \psi  \left(a^4 (2 M+R)+2 a^2 R^2 (3 M+2 R)+3 R^5\right)+R^9 \left(s^2+1\right) \left(-2 M^2
   \left(s^2-2\right)+\right.\right.\right.\\
  & \hspace{40 pt}\left.\left. \left.+M R \left(s^2-4\right)+R^2\right)\right)\right);
\end{align*}

\begin{align*}
& N=-\frac{k^2 M R^2}{\xi  \chi }\left( \left(3 a^6 M R \left(s^2-1\right)^2+a^4 R^2 \left(-2 M^2 \left(2 s^4-7 s^2+3\right)+M R
   \left(7 s^4-29 s^2+12\right)+3 R^2 \left(s^2+1\right)\right)-\right.\right.\\
  & \hspace{40 pt}\left.\left.-\psi  \left(2 a^3 \left(M \left(2-4
   s^2\right)+3 R\right)+6 a R^3\right)+a^2 R^4 \left(-6 M^2 \left(2 s^4-7 s^2+3\right)+M R \left(5 s^4-28
   s^2+3\right)+\right.\right.\right.\\
  & \hspace{40 pt}\left.\left.\left.+4 R^2 \left(s^2+1\right)\right)+R^7 \left(s^2+1\right) \left(M
   \left(s^2-2\right)+R\right)\right)\right)
\end{align*}

\begin{align*}
&{\rm with} \\
&\xi=\sqrt{k^2 \left(a^2+R (R-2 M)\right)}; \\
&\eta=a^4 M \left(s^2-1\right)+a^2 R \left(2 M^2+2 M s^2 (R-2 M)-M R+R^2\right)+R^4 \left(M
   \left(s^2-2\right)+R\right); \\
&\psi=\sqrt{R^3 \left(a^2+R (R-2 M)\right) \left(a^2 M \left(s^2-1\right)^2+R \left(-2 M s^2+M+R s^2\right)
   \left(M \left(s^2-2\right)+R\right)\right)}; \\
&\chi=a^4 M R^2 \left(s^2 (5 M+R)-3 (M+R)\right)+\psi  \left(2 a^3 M+6 a M R^2\right)+a^2 R^3 \left(M^3
   \left(6-12 s^2\right)+3 M^2 \left(2 R s^2+R\right)+\right.\\
  & \hspace{40 pt}\left.+2 M R^2 \left(s^2-3\right)+R^3\right)+R^6 (-(3 M-R))
   \left(M \left(s^2-2\right)+R\right); \\
&\zeta=a^5 R^2 (R-M)+2 a^3 R^3 \left(M^2-3 M R+2 R^2\right)-\psi
   \left(a^4+2 a^2 R (R-2 M)+R^4\right)+3 a R^5 \left(2 M^2-3 M
   R+R^2\right); \\
&\Omega=\frac{\sqrt{\beta ^2-4 \eta  \left(a^2 M s^2-a^2 M-4 M^2 R s^2+2 M^2 R+4 M R^2 s^2-M R^2+R^3
   \left(-s^2\right)\right)}+\beta }{2 \eta }; \\
& \beta=2 a M \left(a^2 \left(s^2-1\right)+R \left(s^2 (3 R-4 M)+2 M-R\right)\right).
\end{align*}

\acknowledgments{}

The work is supported by the Research Centre for Theoretical Physics and Astrophysics, Institute of Physics, Silesian University in Opava, and the GA{\v C}R~\mbox{23-07043S} project.

\end{document}